\documentclass[preprint,11pt,authoryear,nopreprintline]{elsarticle}

\usepackage{amsthm}
\usepackage{amsmath}
\usepackage{booktabs}
\usepackage{listings}
\usepackage{xcolor}

\usepackage{stan}

\definecolor{mauve}{HTML}{B374FD}
\definecolor{codegray}{rgb}{0.5,0.5,0.5}
\definecolor{codepurple}{rgb}{0.58,0,0.82}
\definecolor{backcolour}{rgb}{0.95,0.95,0.92}

\definecolor{function_color}{HTML}{dcbdfb}
\definecolor{keyword_color}{HTML}{65cabe}
\definecolor{comment_color}{HTML}{768390}
\definecolor{entity_color}{HTML}{8ddb8c}

\lstdefinestyle{mystyle}{
    backgroundcolor=\color{backcolour},   
    commentstyle=\color{comment_color},
    keywordstyle=[1]{\color{blue}},
    keywordstyle=[2]{\color{codepurple}},
    keywordstyle=[3]{\color{blue}},
    numberstyle=\tiny\color{codegray},
    stringstyle=\color{codepurple},
    basicstyle=\ttfamily\footnotesize,
    breakatwhitespace=false,         
    breaklines=true,                 
    captionpos=b,                    
    keepspaces=true,                 
    numbers=left,                    
    numbersep=5pt,                  
    showspaces=false,                
    showstringspaces=false,
    showtabs=false,                  
    tabsize=2
}

\usepackage[margin=2.5cm]{geometry}
\usepackage{natbib}
\setcitestyle{numbers,square,super} 

\usepackage{graphicx}
\usepackage{graphicx,bm,hyperref,amssymb,amsmath,amsthm}
\hypersetup{
    colorlinks=true,
    linkcolor=blue,
    filecolor=magenta,      
    urlcolor=cyan,
    pdftitle={Overleaf Example},
    citecolor=cyan,
    pdfpagemode=FullScreen,
}

\usepackage{algorithmic,xcolor, float, url, graphicx}
\usepackage{graphicx}
\usepackage[caption=false]{subfig}

\usepackage{tikz}

\usepackage[title]{appendix}
\usepackage{sourcecodepro}
\usepackage{caption}
\usepackage[lined,ruled,linesnumbered,commentsnumbered]{algorithm2e}
\usetikzlibrary{arrows}

\begin{document}
\begin{frontmatter}
\title{A Short Note on a Flexible Cholesky Parameterization of Correlation Matrices}

\author[1]{Sean Pinkney\corref{cor1}}
\address[1]{Center of Excellence, Omnicom Media Group, New York City}
\ead{sean.pinkney@gmail.com}
\cortext[cor1]{Corresponding author:}

\begin{abstract}
We propose a Cholesky factor parameterization of correlation matrices that facilitates \textit{a priori} restrictions on the correlation matrix. It is a smooth and differentiable transform that allows additional boundary constraints on the correlation values. Our particular motivation is random sampling under positivity constraints on the space of correlation matrices using MCMC methods. 
\end{abstract}

\begin{keyword}
Correlation matrix, Cholesky decomposition, Hamiltonian Monte Carlo, Stan
\end{keyword}
\end{frontmatter}

\section{Introduction}

The $n \times n$ positive-definite correlation matrix plays an important role in many statistical models but is often challenging to estimate, especially in low data environments or complex time varying models. Solutions to the estimation problem trade flexibility - the unrestricted estimation of the matrix - for feasibility by regularization or placing additional restrictions on the matrix   \cite{DEALMEIDA201845}  \cite{Epskamp2018} \cite{LEDOIT2003603} \cite{pourahmadi}. The solution proposed in this note allows researchers an additional way to incorporate structure and known values on some or all of the correlation values.

We propose a parameterization of Cholesky factor of correlation matrices that incorporates user-supplied constraints on the (Pearson) correlation values. The challenge is that the space of correlation matrices is already highly restricted due to the positive definite requirement. The transform sequentially maps unconstrained parameters to the constrained space by enforcing the bounds at each step. This ensures that both the bounds and the positive definiteness of the matrix are satisfied.

The parameterization is defined as a smooth one-to-one transformation from $\mathbb{R}^{n(n-1)/2}$ to the set of $n \times n$ nonsingular Cholesky factors of correlation matrices. It is inspired by the so-called Cholesky–Banachiewicz and Cholesky–Crout sequential Cholesky factorization algorithms of a positive-definite matrix \cite{cholesky} \cite{numerical_recipes}. This method also shows similarities to parameterization methods that use the constrained space to elicit the natural boundaries of correlations \cite{valid4} \cite{lkj} \cite{madar} \cite{numpacharoen2012generating}. These methods, however, are either restricted to lower dimensional correlation matrices or focused on generating correlation matrices without further restrictions. 

Our main contribution is showing how the standard Cholesky algorithm motivates a smooth and differentiable transform that allows additional boundary constraints. These constraints enable many forms of structure such as block structure with equal correlations within blocks, known zeros, or restrictions that all the correlations adhere to a bound such as a positive only requirement.

\section{Method}

The Cholesky–Banachiewicz algorithm factors a known positive-definite correlation matrix $\mathbf{C}$ into a lower Cholesky factor as:

\begin{align} 
L_{j,j} &= \sqrt{1 - \sum_{k=1}^{j-1} L^2_{j,k}} \nonumber \\
L_{i,j} &= \frac{1}{L_{j,j}} \left( C_{i,j} - \sum_{k=1}^{j-1} L_{i,k}L_{j,k} \right) \quad \text{for } i>j.  \label{eq:1}
\end{align}

The value $C_{i,j} \in \mathbf{C}$ is a correlation value that is between $-1, 1$. Expanding the previous formula into matrix form,
\begin{align*}
\mathbf{L} = 
\begin{pmatrix} 
1 &  0 & 0  & \cdots & 0 \\ 
C_{2,1} & \sqrt{1 - L_{2,1}^2} & 0 & \cdots & 0 \\
C_{3,1} &  \left( C_{3,2} - L_{3,1}L_{2,1} \right) /L_{2,2}  &\sqrt{1 - L_{3,1}^2 - L_{3,2}^2} & \cdots & 0 \\
\vdots & \vdots & \vdots & \cdots & \vdots \\
C_{n,1} & \left( C_{n,2} - L_{n,1}L_{2,1} \right) / L_{2,2} & \left( C_{n, 3} - L_{n,1} L_{2,1} - L_{n,2} L_{3,2} \right) / L_{3,3} & \cdots & \sqrt{1 - \sum_{k=1}^{n-1} L_{n,k}}
\end{pmatrix}.
\end{align*}

The entries in the first column are equal to the correlation values in the first column. All the row vectors in the lower triangle of $\mathbf{L}$ are unit-vectors where the diagonal is all positive elements, and, as a result, all the elements must be between ${-1, 1}$ and must be smaller than the square root of the remainder of the length of the unit-vector. Consequently, the boundaries of the Cholesky values are

\begin{align}
\max\left\{ -1 , -\sqrt{1 - \sum_{k=1}^{j-1} L^2_{i,k}}\right\} < L_{i,j} < \min\left\{1, \sqrt{1 - \sum_{k=1}^{j-1} L^2_{i,k}} \right\}.  \label{eq:2}
\end{align}

For the purposes of generating differentiable transforms we exclude $\pm 1$ correlations and make the inequalities strict. We wish to place boundaries on each $C_{i,j}$ in $\mathbf{C}$ such that $-1 \le a_{i,j} < C_{i,j} < b_{i,j} \le 1$. To accomplish this we will multiply the previous inequality by $|L_{j,j}|$ and add $\sum_{k=1}^{j-1} L_{i,k}L_{j,k}$. In fact, this is the same as solving for $C_{i,j}$ in equation \ref{eq:1}:

\begin{align}
-L_{j,j} + \sum_{k=1}^{j-1} L_{i,k}L_{j,k} \le a_{i,j} < C_{i,j} < b_{i,j} \le L_{j,j} + \sum_{k=1}^{j-1} L_{i,k}L_{j,k} 
\label{eq:3}
\end{align}
where the absolute value is expanded into $-L_{j,j}$ for the lower bound and $L_{j,j}$ for the upper bound. The correlation value, $C_{i,j}$, must then fall between the solved for bounds.

Re-arranging equation \eqref{eq:3} we can express this in terms of the elements of the lower triangular Cholesky factor. There are two bounds which need to be satisfied. The first bound of the entry in the Cholesky factor is the bound implied by the bounds on the values, $a, b$. This bound is found by placing $a, b$ into equation \eqref{eq:1} for the correlation value $C$. The second bound is found by incorporating the unit-vector restriction from \eqref{eq:2} for the $i^{\text{th}}$ row. This is given in equation \eqref{eq:2}. With these two conditions and the given correlation bounds, $a, b$, the $L_{i,j}$ entry of the Cholesky factor is bounded by

\begin{align}  \label{eq:4}
-\sqrt{1 - \sum_{k=1}^{j-1} L^2_{i,k}} \; \le \; \frac{a_{i,j} - \sum\limits_{k=1}^{j-1} L_{i,k}L_{j,k}}{L_{j,j}} < \; & L_{i,j} \;  < \frac{b_{i,j} - \sum\limits_{k=1}^{j-1} L_{i,k}L_{j,k}}{L_{j,j}} \; \le \;\sqrt{1 - \sum_{k=1}^{j-1} L^2_{i,k}} \\
\max\left\{ -\sqrt{1 - \sum_{k=1}^{j-1} L^2_{j,k}} ,\; \frac{a_{i,j} - \sum\limits_{k=1}^{j-1} L_{i,k}L_{j,k} }{L_{j,j}} \right\} < &\; L_{i,j} \; < \min\left\{\sqrt{1 - \sum_{k=1}^{j-1} L^2_{j,k}}, \; \frac{b_{i,j} - \sum\limits_{k=1}^{j-1} L_{i,k}L_{j,k}}{L_{j,j}} \right\},
\end{align}
where $i > j$ and $i \in (3, \ldots, n)$. The first column of the lower triangular Cholesky factor is mapped from $\mathbb{R}$ to the bounds by a bijective transform $f(x_{i,1} \mid a_{i,1}, b_{i,1})$. The next $n - 2$ columns use the above inequality to ensure that the Cholesky factor elements adhere to the given bounds. Lastly, the diagonal elements are calculated by the remaining length of the unit vector.

\subsection{Jacobian Determinant}

The inverse transform function from the mapping of each $\mathbb{R}$ to $(a, b)$ is $f(x) \in (a, b)$ and $-1 \le a < b \le 1$. We choose the scaled and shifted inverse logit function defined as 
$$
f(x \mid a, b) = a + \frac{b - a}{1 + e^{-x}}.
$$
The absolute Jacobian determinant of the inverse transform is the absolute derivative of a one variable function
$$
g(x) = \frac{d}{dx}f(x) = (b - a)f(x)(1 - f(x)).
$$
The function is positive since $b -a > 0$ and $f(x) > 0$. As each $L_{i,j}$ is mapped directly using this function, the derivative of this is the Jacobian determinant.

\subsection{Algorithm}

The algorithm for the mapping is given in Algorithm \ref{alg:chol}. The input bounds, $a$ and $b$, may be scalars or vectors. The Jacobian determinant is calculated in $\left|J\right|$. The input vector, $x$, is a random vector of values from the real line. The functions $f(\cdot)$ and $g(\cdot)$ refer to the scaled and shifted inverse logit function and its derivative. 

\begin{algorithm}[H] 
\caption{Cholesky Parameterization Algorithm} \label{alg:chol}
\KwIn{$x$ a real vector of length $n (n - 1) / 2$} 
\KwData{$a$ and $b$ for the lower and upper bounds of each value} 
\KwResult{$\mathbf{L}$ a lower triangular Cholesky factor of a correlation matrix}
$L_{1,1} \gets 1$ \\
$L_{2:n,1} \gets f(x[1:n - 1], a, b)$ \\
$L_{2,2} \gets \sqrt{1 - L_{2,1}^2}$ \\
$\left | J \right| = \prod_{w=1}^{n-1} g(x[w], a, b)$ \\

\For{$i \gets 3$ \KwTo $n$ }{
  $y = \sqrt{1 - L_{i,1}^2}$ \\
    \For{$j \gets 2$ \KwTo $i - 1$ } {
     $z = L_{j, 1:j - 1}^T L_{i, 1:j - 1}$  \\
     $lb = \max \left(-y, \frac{a - z}{L_{j, j}} \right)$ \\ 
     $ub = \min \left(y, \frac{b - z}{L_{j, j}} \right)$ \\
     $L_{i,j} \gets f(x[\cdot], lb, ub)$ \\
     $\left | J \right | \mathrel{\raisebox{0.19ex}{$\scriptstyle*=$}} g(x[\cdot], lb, ub)$ \\
     $y \mathrel{\raisebox{0.19ex}{$\scriptstyle*=$}} \sqrt{1 - (L_{i, j} / y)^2}$
      }
    $L_{i,i} \gets y$
}
\end{algorithm} 

\subsection{Impossible Bounds} \label{bound_refs}

The user defined bounds must be chosen carefully as certain valid values for a particular correlations result in downstream correlations violating the boundary region. 

For example, with a $3 \times 3$ matrix that we wish to constrain all correlation values to be negative. As noted above the first column of the Cholesky factor is allowed to span the full region because the length of the unit-vector is maximal and only the given bounds must be satisfied. Let us choose $C_{2, 1} = C_{3,1} = \frac{-1}{\sqrt{2}}$ and solve for the bounds of $C_{3,2}$:

$$
\begin{aligned}
\max\left\{-1, -\sqrt{1 - C_{3,1}^2} \right\} &< \frac{B - C_{2,1} C_{3,1}}{L_{2, 2}} < \min\left\{0, \sqrt{1 - C_{3,1}^2} \right\} \\
-\sqrt{0.5} &< \frac{B - 0.5}{\sqrt{0.5}} < 0 \\
\implies 0 &< B < 0.5.
\end{aligned}
$$
The boundary condition must be greater than or equal to zero violating our initial condition that all correlations are negative. 

When using this transform with a derivative based sampler such as Hamiltonian Monte Carlo, violation of the bounds will result in discontinuities and the sampler will fail. We suggest giving bounds as data into the program and regularizing large correlations with a prior that restricts large correlation values such as that with an LKJ. 

\section{Discussion}

This short note shows a new method to sample Cholesky factors of correlation matrices with additional boundary constraints.  

\subsubsection*{Acknowledgements}

We would like to thank Enzo Cerullo for his encouragement to pursue the research into this transform, Seth Axen, Benjamin Goodrich, and Stephen Martin for their comments and feedback. All errors are solely due to the author. 

\bibliography{all}{}

\begin{appendices}

\section{Stan Code}

Below is a sample Stan program for this parameterization. The bounds are input as scalar values for all the correlation entries to be within. This can easily be generalized to incorporate different bounds for each correlation value. 

One obvious change is to set certain correlation values. It is possible to set a given correlation value to a value though values around zero will sample much easier for the reason of violating constraints in \ref{bound_refs}. 

To add known correlation values, place all the known correlation values that into an $m$-length vector $\mathbf{p}$ where $m < n(n-1)/2$. Only the unknown $n(n-1)/2 - m$ unconstrained parameters remain for the parameterization. The Cholesky factor entry for the known values will be 

$$
L_{i, j} \stackrel{set}{=} \frac{p_{m} - \sum_{k=1}^{j-1} L_{i,k}L_{j,k}}{L_{j, j}}.
$$

\begin{lstlisting}[language=Stan, caption=Stan Example]
functions {
  vector lb_ub_lp (vector y, real lb, real ub) {
    target += log(ub - lb) + log_inv_logit(y) + log1m_inv_logit(y);
    
    return lb + (ub - lb) * inv_logit(y);
  }
  
  real lb_ub_lp (real y, real lb, real ub) {
    target += log(ub - lb) + log_inv_logit(y) + log1m_inv_logit(y);
    
    return lb + (ub - lb) * inv_logit(y);
  }
  
  matrix cholesky_corr_constrain_lp (vector col_one_raw, vector off_raw,
                                           real lb, real ub) {
    int K = num_elements(col_one_raw) + 1;
    vector[K - 1] z = lb_ub_lp(col_one_raw, lb, ub);
    matrix[K, K] L = rep_matrix(0, K, K);
    L[1, 1] = 1;
    L[2:K, 1] = z[1:K - 1];
    L[2, 2] = sqrt(1 - L[2, 1]^2);
    
    int cnt = 1;
    
    for (i in 3:K) {
       real l_ij_old = sqrt(1 - L[i, 1]^2);
       L[i, 2] = l_ij_old;
      for (j in 2:i - 1) {
        real stick_length_x_l_jj = l_ij_old * L[j, j];
        real b1 = dot_product(L[j, 1:(j - 1)], L[i, 1:(j - 1)]);
          
          // how to derive the bounds
          // we know that the correlation value C is bound by
          // b1 - Ljj * Lij_old <= C <= b1 + Ljj * Lij_old
          // Now we want our bounds to be enforced too so
          // max(lb, b1 - Ljj * Lij_old) <= C <= min(ub, b1 + Ljj * Lij_old)
          // We have the Lij_new = (C - b1) / Ljj
          // To get the bounds on Lij_new is
          // (bound - b1) / Ljj 
          
          real low = max({-stick_length_x_l_jj, lb - b1});
          real up = min({stick_length_x_l_jj, ub - b1});
          real x = lb_ub_lp(off_raw[cnt], low, up);
          
          // dividing by L[j, j] differs from algo in paper 
          // In the paper, the division is handled inside the bounds so
          // low <- low / L[j, j] 
          // up <- up / L[j, j] 
          // but it seems to be more numerically stable to do it after
          // and adjust with the -log(L[j,j]) for the division
          
          L[i, j] = x / L[j, j]; 
          target += -log(L[j, j]);
          
          l_ij_old *= sqrt(1 - (L[i, j] / l_ij_old)^ 2);
          cnt += 1;
        }
         L[i, i] = l_ij_old;
      }
        return L;
  }
 }
 data {
  int<lower=2> K; // dimension of correlation matrix
  real<lower=0> eta;
  real<lower=-1> lb;
  real<upper=1> ub;
 }
 transformed data {
  int k_choose_2 = (K * (K - 1)) %/% 2;
  int km1_choose_2 = ((K - 1) * (K - 2)) %/% 2;
 }
 parameters {
  vector[K - 1] col_one_raw;
  vector[km1_choose_2] off_raw;
 }
 transformed parameters {
  matrix[K, K] L_Omega = cholesky_corr_constrain_lp(col_one_raw, off_raw, lb, ub);
 }
 model {
  L_Omega ~ lkj_corr_cholesky(eta);
 }
 generated quantities {
  matrix[K, K] Omega = multiply_lower_tri_self_transpose(L_Omega);
 }
\end{lstlisting}

\end{appendices}
\end{document}